# Evidence for short-range magnetic order in the nematic phase of FeSe from anisotropic in-plane magnetostriction and susceptibility measurements

Mingquan He,[1, ∗] Liran Wang[∗],[1, †] Frédéric Hardy,[1] Liping Xu,[1, 2]
Thomas Wolf,[1] Peter Adelmann,[1] and Christoph Meingast[1, ‡]

[1]*Institute for Solid State Physics, Karlsruhe Institute of Technology, 76021 Karlsruhe, Germany*
[2]*Key Laboratory of Polar Materials and Devices,
Ministry of Education, Department of Electronic Engineering,
East China Normal University, Shanghai 200241, China*
(Dated: 12/09/17)

The nature of the nematic state in FeSe remains one of the major unsolved mysteries in Fe-based superconductors. Both spin and orbital physics have been invoked to explain the origin of this phase. Here we present experimental evidence for frustrated, short-range magnetic order, as suggested by several recent theoretical works, in the nematic state of FeSe. We use a combination of magnetostriction, susceptibility and resistivity measurements to probe the in-plane anisotropies of the nematic state and its associated fluctuations. Despite the absence of long-range magnetic order in FeSe, we observe a sizable in-plane magnetic susceptibility anisotropy, which is responsible for the field-induced in-plane distortion inferred from magnetostriction measurements. Further we demonstrate that all three anisotropies in FeSe are very similar to those of $BaFe_2As_2$, which strongly suggests that the nematic phase in FeSe is also of magnetic origin.

Magnetism appears to be the universal driving force for high-temperature superconductivity in e.g. cuprates and Fe-based compounds[1, 2]. However, this scenario has been challenged by the structurally simple iron chalcogenide FeSe. Unlike iron-pnictide compounds, long-range magnetic order is absent in stoichiometric FeSe at ambient pressure, although it does undergo a similar structural transition to an electronic nematic state [3–6]. The microscopic nature of this state of reduced rotational symmetry, from which superconductivity emerges, remains enigmatic, and both spin [7–10] and orbital [11–14] degrees of freedom have been intensively discussed. At first glance, the absence of static magnetism seems to discredit the spin-nematic scenario and favors an orbital order [15–19]. However, recent theoretical proposals indicate that the magnetic interactions in FeSe are highly frustrated, suppressing magnetic (but not nematic) order [20–22]. Experimentally, this interpretation is supported by the observation of low-energy spin fluctuations along the $(\pi,0)$ wave-vector below the nematic transition at $T_S$ [23–25]. To date, the nematic phase of FeSe has been studied by means of elastic modulus [15, 26], transport [6, 27], inelastic neutron [23–25] and Raman spectroscopies [28], ARPES [6] and NMR measurements [15–17]. Direct measurements of the in-plane magnetic anisotropy like in $BaFe_2As_2$, which allows to disentangle between magnetic and orbital orders [29], are, however, still lacking.

In this Letter, the anisotropic magnetic response of FeSe is studied using a combination of magnetostriction, magnetic susceptibility and resistivity measurements on FeSe single crystals in order to unravel the nature of the nematic state. Magnetization measurements on uniaxially strained FeSe clearly show a substantial in-plane magnetic susceptibility anisotropy developing within the nematic phase. This anisotropy agrees well with our magnetostriction measurements, which provide an indirect measure of the susceptibility anisotropy. Surprisingly, the temperature dependence of both the susceptibility and transport anisotropies are extremely similar to that of long-range magnetically ordered $BaFe_2As_2$, although the signs of both quantities are reversed. It was theoretically demonstrated that orbital order alone is insufficient to produce a sizable susceptibility anisotropy [29] and that magnetic order and spin-orbit coupling are essential. Here using this same reasoning, we argue that our data therefore provide strong evidence for short-range magnetic order in the nematic phase of FeSe, as has been suggested in several theoretical works [20–22].

Vapor-grown single crystals of FeSe [5, 15, 30], with typical dimensions of roughly 2 mm × 2 mm × (0.06 - 0.2) mm, were selected for this study. The high $T_c$ = 9.1 K determined by heat capacity (see supplemental material Fig. S1[31]) and resistivity [see Fig. 2(b)], large residual resistivity ratio RRR=R(300 K)/R(0 K)∼166 [see Fig.2(b)] both demonstrate the high quality of our single crystals. Resistivity anisotropy measurements using the glass-fiber-reinforced plastic (GFRP) substrate method, which was previously successfully employed for anisotropically straining $BaFe_2As_2$[29], proved to be ineffective for applying a large strain to FeSe, i.e. no significant resistivity anisotropy could be observed. Most likely the crystals exfoliate due to the weak interlayer bonding using this method, and the top layer with the electrical contacts remains unstrained. We therefore used a 'gentler' method, in which two opposing ends of the crystal along the $[110]_{tet}$ direction are glued to a polyether ether ketone (PEEK) substrate using GE varnish [see Fig.2(a) inset]. The PEEK material has a much larger thermal-expansion coefficient than FeSe (along the a-axis) and thus exerts a positive uniaxial strain on FeSe upon cooling (see supplemental material Fig. S2 [31]). As a result, clear anisotropies in both resistivity and magnetization could be observed using this uniaxial straining method. Resistivity measurements were made using a standard four-terminal geometry. Thermal expansion and magnetostriction measurements were made using a home-built high-resolution capacitance dilatometer [32]. Magnetization measurements were carried out in a Physical Property Measurement System (PPMS) using the Vibrating-Sample-Magnetometer (VSM) from Quantum Design.

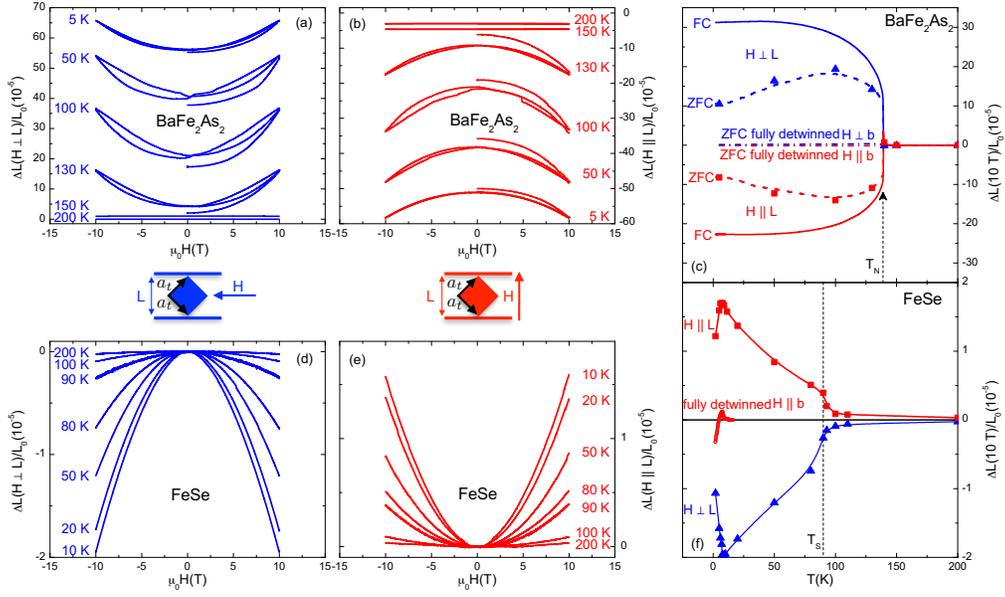

FIG. 1. Magnetostriction along the $[110]_{tet}$ direction of Ba122 (a-b) and FeSe (d-e) 'twinned' single crystals for magnetic fields applied parallel and perpendicular to the length $L(H,T)$, as indicated by the accompanying illustrations. Field cooled (FC) and zero-field cooled (ZFC) differences in length changes between 0 T and 10 T, derived from thermal expansion measurements for $BaFe_2As_2$ (c) and taken directly from (d-e) for FeSe. As outlined in the text, a significant magnetostriction results from a field-induced change in twin population resulting from a magnetic anisotropy in the orthorhombic phase. This magnetostriction practically vanishes for fully detwinned crystals [see (c, f)].

It has previously been shown that a high magnetic field applied along the $[110]_{tet}$ direction of the original tetragonal cell can be used to detwin $BaFe_2As_2$ crystals [33–35]. This field detwinning was attributed to an in-plane anisotropy of the magnetic susceptibility in the ordered state below $T_{S,N}$ [33, 34], and, due to the considerable orthorhombic distortion within each magnetic domain, high-resolution magnetostriction measurements are expected to provide a very sensitive method for studying this effect. To set the stage, we first present magnetostriction data on $BaFe_2As_2$, which has long-range magnetic order, and will then compare these data to those of FeSe.

To study field-induced detwinning, one needs to use a thick $BaFe_2As_2$ single crystal, for which the small force applied by the dilatometer is not sufficient to detwin [36, 37] the sample. The magnetostriction of such a 'thick crystal' are shown in Figs. 1a and 1b for two different field orientations and for temperatures between 200 K and 5 K. Significant magnetostriction is only observed below $T_{S,N}$=139 K. For the configuration $H\perp L$, $L$ increases with field, whereas $L$ decreases with field for $H\|L$. All curves below $T_{S,N}$ exhibit a considerable hysteresis, which can be attributed to pinning of domain walls. We also performed field-cooled (FC) and zero-field-cooled (ZFC) thermal-expansion measurements in 10 T, from which the magnetostriction at 10 T was determined by subtracting the zero-field data as shown in Fig.1(c). These data clearly show that the field-induced detwinning process starts abruptly below $T_N$. The solid symbols in Fig.1(c) are taken from Figs.1(a)(b) and match the ZFC data very well. For fully detwinned crystals the magnetostriction practically vanishes as shown by the dash-dotted lines in Fig.1(c). This demonstrates that the observed magnetostriction is due to field-induced detwinning and not due to an intrinsic magnetostriction of the stripe magnetic state. The sign of the magnetostriction suggest that the shorter $b$-axis has the higher susceptibility, which is consistent with direct measurements [29]. Domains with the $b$-axis aligned along the field direction expand in population to lower the energy, as a result, $L$ decreases(increases) with field for $H\|L$(for $H\perp L$) as shown in Figs.1(a)(b).

In order to probe the nematic state in FeSe, we performed the same magnetostriction measurements on a 'thick twinned' FeSe crystal [see Figs.1(d-f)]. The overall behavior is remarkably similar to that for $BaFe_2As_2$; i.e.

the magnitude of the magnetostriction increases abruptly below $T_S$ and is negligible above $T_S$. Similarly, for a 'thin' fully detwinned sample, this magnetostriction signal vanishes [see Figs.1(f)]. There are however also several important differences. First, the sign of the magnetostriction of FeSe is opposite to that of $BaFe_2As_2$, and the magnitude is about 10 times smaller. Further, the magnetostriction of FeSe is free of hysteresis at all temperatures and the temperature dependence is quite different, increasing continuously down to low temperature. The implications of these results will be discussed later. Here we note that, using the above data, we can estimate that roughly 30 T and 100 T are needed to fully detwin $BaFe_2As_2$ and FeSe, respectively (see supplemental material Section IV [31] ), in good agreement with Ref.[34] for $BaFe_2As_2$. Further, the magnetic field can be translated to a uniaxial pressure, and at these respective fields we find a uniaxial pressure of about 10 MPa for $BaFe_2As_2$ and 15 MPa in FeSe, which also agrees well with typical pressures needed to detwin these crystals [38].

Since our magnetostriction data only provide indirect evidence for the susceptibility anisotropy, we also made an effort to measure this anisotropy directly by applying a uniaxial strain using the differential thermal-expansion between FeSe and a PEEK sample holder, as described above. We estimate (see supplemental material Section III [31]) that at the structural transition a strain of about $1 \times 10^{-3}$ (approximately 30% of the spontaneous distortion) can be expected, which is sufficient to observe the in-plane anisotropy of the magnetic susceptibility, but not sufficient to exfoliate the crystal.

Figure 2(a) displays the resulting $a$- and $b$- axis magnetic susceptibilities, for H = 12 T, together with the twinned measurement. No background subtraction is needed in these measurements here, since the long weakly magnetic PEEK sample holder has essentially no signal in the VSM magnetometer. Above $T_S$, no difference between $\chi_a$ and $\chi_b$ can be resolved and the susceptibility scales linearly with temperature, as also observed in iron pnictides [29, 39–41]. A kink around 90 K in both directions signals the nematic/structural transition, below which a clear splitting between $\chi_a$ and $\chi_b$ becomes evident. We find that the susceptibility measured along the shorter $b$-axis is smaller than that of the $a$-axis. This anisotropy is seen more clearly in the lower right inset of Fig. 2(a), in which the difference $\Delta\chi = \chi_b - \chi_a$ is plotted. The susceptibility anisotropy grows continuously from 0 above $T_S$ to low temperature. Our results are quite similar to those of $BaFe_2As_2$[29] in the sense that the anisotropy only develops below $T_S$. Interestingly, we find $\chi_b < \chi_a$ for FeSe which is opposite in sign to that of $BaFe_2As_2$, for which $\chi_b > \chi_a$ within the stripe anti-ferromagnetic(AF) phase[29]. This sign reversal also applies for the resistivity anisotropy, which we also measured using strain applied from a PEEK substrate [see Fig. 3]. The sign of the resistivity anisotropy agrees with previous studies [6, 27], however its magnitude varies greatly between the different measurements, which we attribute to the intrinsic difficulty of applying a well defined strain to FeSe. Details of resistivity measurements of FeSe are given in supplemental material. We note that an anisotropy of the Knight shift starting slightly above $T_S$ in twinned crystals has also been observed in NMR measurements, however the sign of the anisotropy could not be determined due to the twinned nature of the crystals [15–17].

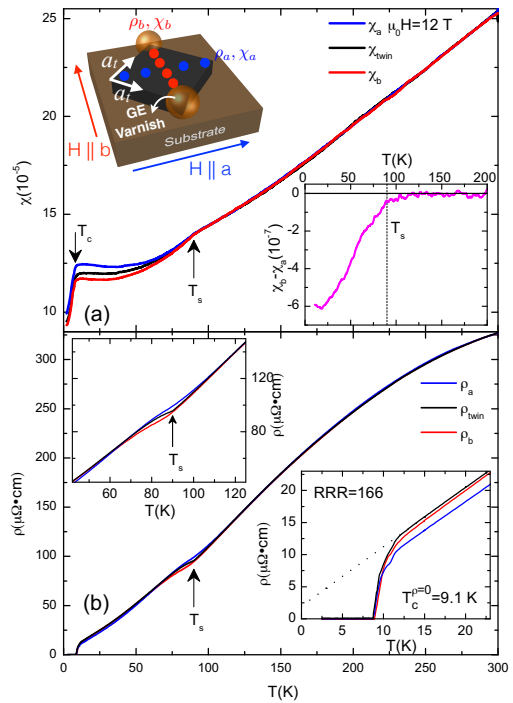

FIG. 2. (a) Temperature dependence of in-plane susceptibility anisotropy of FeSe in a field of 12 T. The crystal was detwinned using the thermally induced strain from the PEEK substrate sample holder (see inset in upper left and text). The black line is the susceptibility in the twinned state, taken on the same crystal. The inset on the lower right shows that the in-plane susceptibility anisotropy $\chi_b - \chi_a$ develops below $T_S$. (b) Resistivity anisotropy of FeSe measured using the same uniaxial strain setup as for the susceptibility measurements. A clear anisotropy is observed close to $T_S$ (see inset in upper left). The resistivity in the twinned state shows a superconducting transition at 9.1 K and has a quite high RRR ratio of about 170.

We now discuss the implication of our experimental results. In Figure 3 we compare the in-plane susceptibility and resistivity anisotropies of FeSe and $BaFe_2As_2$[29]. Except for the opposite signs and different magnitudes of both effects, we find very similar behavior in both systems. Whereas the resistivity anisotropy develops well above and diverges upon approaching $T_S$ or $T_N$ from above, the susceptibility anisotropy only appears below $T_S$ or $T_N$. It was previously demonstrated that orbital order alone is insufficient to produce a sizable suscep-tibility anisotropy and that magnetic order and spin-orbit coupling are essential [29]. Using this same reasoning, we argue that our data therefore provide strong evidence for some kind of magnetic order also in FeSe. Since there exists no evidence for long-range magnetic order in FeSe, likely candidates for magnetism are short-range frustrated magnetic orders in the nematic phase of FeSe, as suggested in several theoretical works [20–22]. We note that the proposed frustrated orders of Ref. [22] locally have a very similar ordering as in the long-ranged ordered stripe phase, forming a kind of phase-disordered AF chains. In a local picture, even such a short-range order is expected to result in a susceptibility anisotropy, albeit with a significantly reduced magnitude. The opposite sign of the susceptibility of FeSe (compared to $BaFe_2As_2$) suggests that the spins, on average, are aligned along the shorter orthorhombic axis [29], in contrast to $BaFe_2As_2$.

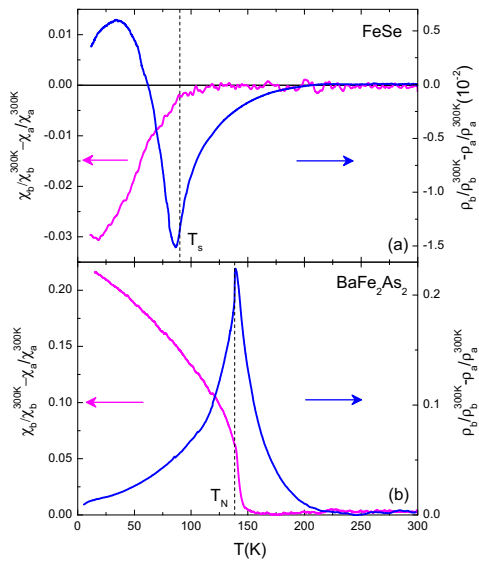

FIG. 3. Comparison of the susceptibility and resistivity anisotropies of (a) FeSe and (b) $BaFe_2As_2$. The data of $BaFe_2As_2$ are taken from Ref. [29]. Except for the magnitude and sign, the behavior of both systems is quite similar.

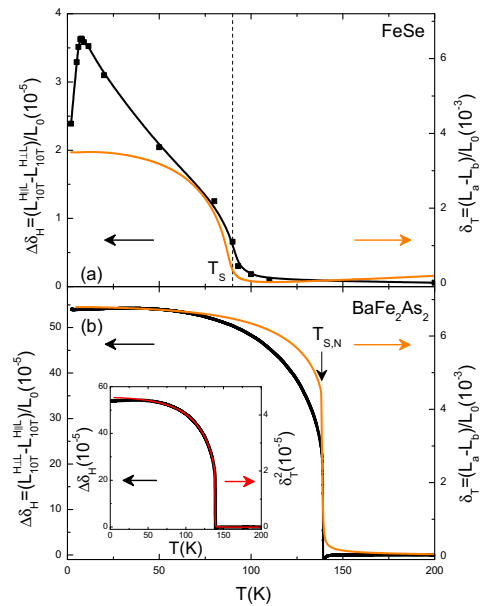

FIG. 4. Attempted scaling of the field-induced distortion at 10 T, $\Delta\delta_H$, with the spontaneous orthorhombic distortion, $\delta_T$, for (a) FeSe and (b) $BaFe_2As_2$. Whereas a rough scaling is observed for $BaFe_2As_2$, this scaling does not work for for FeSe. In particular, $\Delta\delta_H$ of FeSe continues to increase to lower temperatures, whereas $\delta_T$ becomes flat. The inset in (b) shows that nearly perfect scaling can be obtained for $BaFe_2As_2$ if one scales $\Delta\delta_H$ with $\delta_T^2$.

Finally, in Fig. 4, we compare the temperature dependence of the field-induced distortion $\Delta\delta_H$ to the zero-field spontaneous distortion $\delta_T$, inferred from our thermal-expansion data. For $BaFe_2As_2$, both quantities have similar temperature dependences suggesting an intimate connection between the magnetic order and structural distortion. In fact, $\Delta\delta_H$ scales perfectly with $\delta_T^2$ [see inset of Fig. 4(b)], which may provide important details about the magnetostrictive coupling in this material [42]. In contrast, there is no clear relation between these quantities for FeSe. Instead of flattening at low $T$, $\Delta\delta_H$ of FeSe continues to increase down to $T_c$. This suggests that either the strength or the nature of the magnetic short-range order in FeSe is strongly tempera-ture dependent, which is consistent with the frustrated scenario. Indeed, similar to the susceptibility anisotropy, the spin-relaxation rate in FeSe also only emerges below $T_S$ and diverges at low temperature before $T_c$ is approached [15–17]. The frustrated magnetic ground state is moreover strongly supported by the observation of both stripe- and Néel-type spin fluctuations in recent inelastic neutron scattering (INS) experiments [23–25]. Our findings therefore point to a strong involvement of the spin degrees of freedom in the nematic transition of FeSe.

In summary, the nematic phase of FeSe has been studied using measurements of the in-plane anisotropies of the uniform magnetic susceptibility, the magnetostriction and the resistivity. Similar to $BaFe_2As_2$, the susceptibility and magnetostriction anisotropies develop only below the nematic transition temperature, whereas the resistivity anisotropy starts to develop at much higher temperatures. The sizable susceptibility anisotropy in these systems is due to spin-orbit coupling and develops only in the presence of magnetic order [29]. Our results thus strongly support the existence of some kind of short-range magnetic order within the nematic phase of FeSe and suggest that nematicity in iron-based systems is universally induced by magnetism.

We thank Ilya Eremin, Igor Mazin and Qimiao Si for valuable discussions.

# Supplemental Material: Evidence for short-range magnetic order in the nematic phase of FeSe from anisotropic in-plane magnetostriction and susceptibility measurements

Mingquan He,[1] Liran Wang,[1] Frédéric Hardy,[1] Liping Xu,[1,2]
Thomas Wolf,[1] Peter Adelmann,[1] and Christoph Meingast[1,*]

[1] *Institute for Solid State Physics, Karlsruhe Institute of Technology, 76021 Karlsruhe, Germany*
[2] *Key Laboratory of Polar Materials and Devices,
Ministry of Education, Department of Electronic Engineering,
East China Normal University, Shanghai 200241, China*

## I. SAMPLE QUALITY

Figure S1 shows the resistivity and specific heat of a typical FeSe sample used in the magnetostriction and magnetization measurements. The discontinuous jump in the specific heat data implies that bulk superconducting transition occurs at $T_c = 9.1$ K [Fig.S1(b)]. It is the same temperature below which the resistivity becomes zero as shown in Fig.S1(a). The residual resistivity ratio (RRR) is estimated based on a linear extrapolation of the normal state resistivity to zero temperature, we obtain RRR=R(300 K)/R(0 K)∼166. Large values of $T_c$ and RRR prove that our samples are of high quality.

## II. DETWINNING SETUP

The detwinning setup is shown in Fig.S2(a). The the electrical contacts for resistivity measurements are also illustrated. Single crystals were glued on top of a Polyether Ether Ketone (PEEK) substrate by fixing two $[110]_t$ ends with GE varnish. For the magnetization measurements, a home made PEEK sample holder was used, which has negligible magnetic response. Figure S2(b) shows the temperature dependence of the thermal-expansion of the PEEK substrate and a free standing FeSe sample along the two orthorhombic axes ($a > b$). The thermal-expansion of PEEK is much larger than that of FeSe, due to the differential thermal expansion, uniaxial symmetry breaking strain is therefore applied to the sample which effectively detwins the crystal when the whole detwinning setup is cooled down.

## III. STRAIN ESTIMATION

In order to estimate the uniaxial strain applied to the FeSe crystal in our detwinning method, we performed in-plane resistivity anisotropy measurements and compare the corresponding elastoresistivity with that obtained by piezoelectric stack measurements [1]. To eliminate the uncertainty of different strain conditions from sample to sample, the resistivity along both orthorhombic axes was measured on the same sample simultaneously.

The measured resistivity of twinned and detwinned

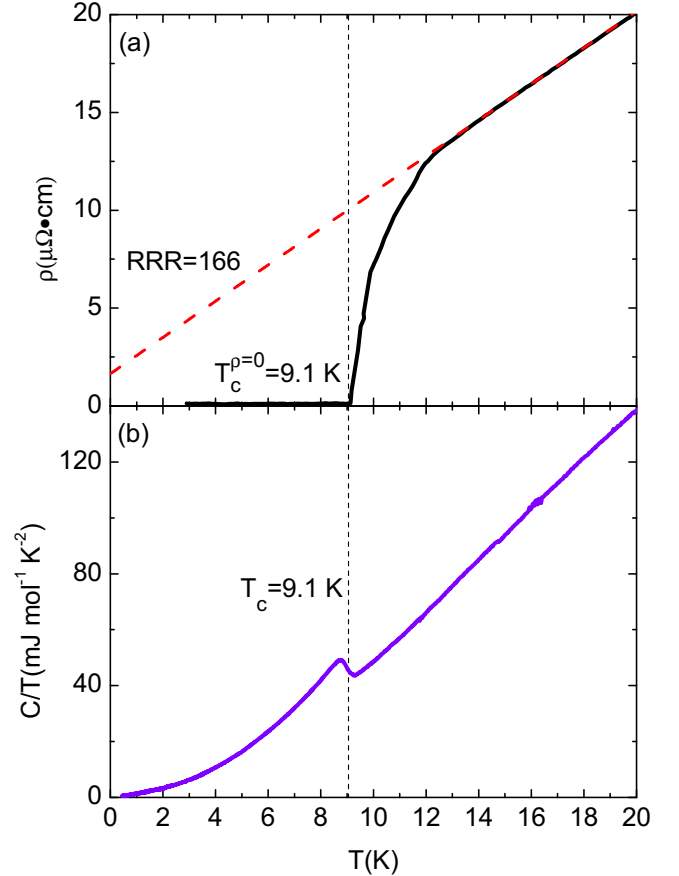

FIG. S1. Temperature dependence of the (a) resistivity and (b) specific heat in the vicinity of the superconducting transition. Both data shows that $T_c = 9.1$ K. Red dash line in (a) is an extrapolation according to a linear fit of the resistivity from 15 K to 30 K, which gives the residual resistivity ratio RRR=R(300 K)/R(0 K)∼166.

FeSe sample is displayed in Fig. S3(a). Unlike the pronounced resistivity anisotropy observed in the $BaFe_2As_2$ system, the difference between $\rho_a$ and $\rho_b$ can barely be seen well above $T_S$. A closer look indicates that the anisotropy develops gradually below 200 K where the resistivity along the shorter axis $\rho_b$ becomes smaller than that of the longer axis $\rho_a$ until $T^*=62$ K below which $\rho_b > \rho_a$ (see Fig. S4). The anisotropy above (below) 62 K has the same (opposite) sign as FeTe [2] and hole doped BaFe$_2$As$_2$ [3], but has reversed (identical) sign as undoped and electron underdoped BaFe$_2$As$_2$ [4–8]. As shown in the inset of Fig. S3(a), the structural transition of the strained state is broadened in comparison with the twinned case. This broadening of the structural transition is seen more clearly in the temperature derivative $d\rho/dT$ shown in Fig. S3(b). In the twinned state, a very sharp jump occurs at $T_S = 90$ K in $d\rho_T/dT$, which turns into a rather broad dip $d\rho_{a,b}/dT$ under strain. In addition to the smeared out transition, the transition temperature is also pushed down about 1.5 K, which is the consequence of negative pressure dependence of the structure transition [9, 10]. This implies that the strain is successfully transmitted to the sample.

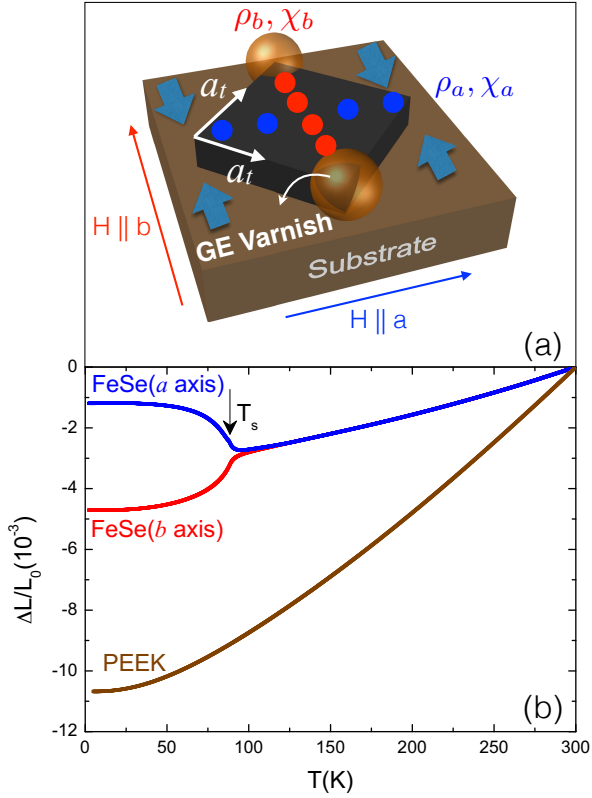

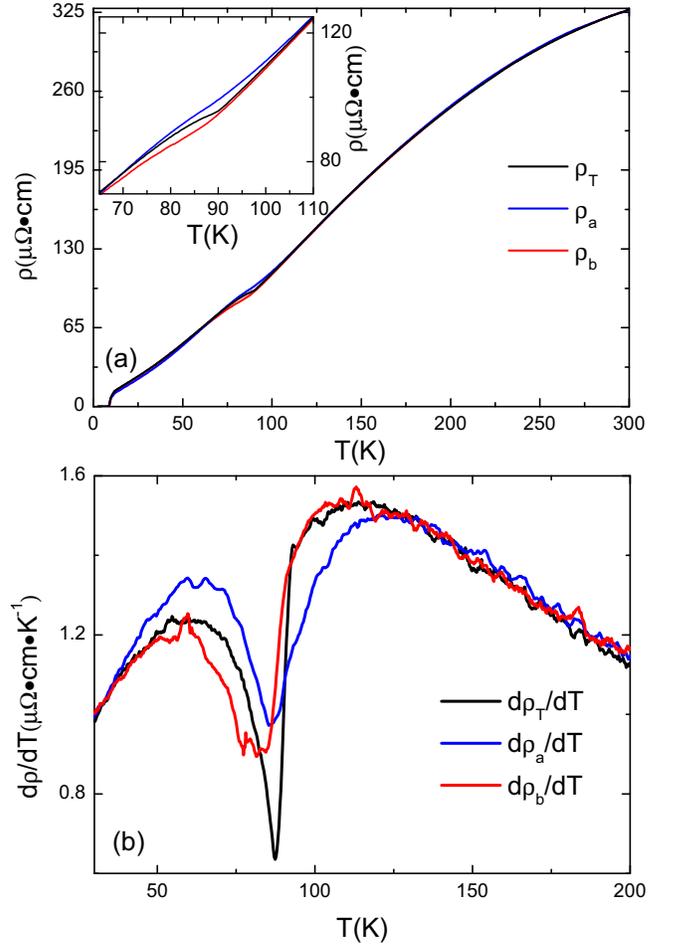

FIG. S2. (a) Schematic model of the uniaxial detwinning device. Blue and red dots are the electrical contacts for resistivity measurements along $a$ and $b$ axes respectively. (b) Thermal expansion of the PEEK substrate in comparision with a free standing FeSe sample along orthorhombic $a$ and $b$ axes.

FIG. S3. (a) Temperature dependence of resistivity in detwinned and twinned FeSe. The inset in (a) shows enlarged view near $T_S$ where the structure transitions are broadened in both directions. (b) Temperature derivative of the resistivity showing that the structural transition is broadened under strain.

To gain a quantitative picture of the transmitted strain, using the resistivity data shown in Fig. S3, we adjust the strain to match the $m_{66}$ channel of the elastoresistivity probed by piezoelectric stack measurements[1]:

$$2m_{66}(T) = \frac{\rho_b(T) - \rho_a(T)}{\rho_T(T)\left(\varepsilon_b(T) - \varepsilon_a(T)\right)}, \quad (1)$$

$$\varepsilon_b(T) - \varepsilon_a(T) = \varepsilon_b(T)(1 + \nu), \quad (2)$$

$$\varepsilon_b(T) \approx \gamma \cdot [\Delta L_{PEEK}/L_0 - \Delta L^{twin}_{FeSe}/L_0], \quad (3)$$

where $\upsilon$ is the Poisson's ratio of FeSe extracted from ultrasound experiments [11], $\gamma \sim 5\%$ is the strain transmission coefficient which is estimated by scaling our data with the $2m_{66}$ obtained by piezoelectric stack measurements [1]. The calculated $2m_{66}$ are presented in Fig. S4(a) and the same convention of $-2m_{66}$ is plotted in order to compare with BaFe$_2$As$_2$. The obtained $2m_{66}$ scales excellently with piezoelectric stack experiments above the transition, which exhibits a divergent Curie-Weiss behavior approaching $T_S$ from above. An unusual sign change occurs at $T^*$=62 K which is higher than that observed by

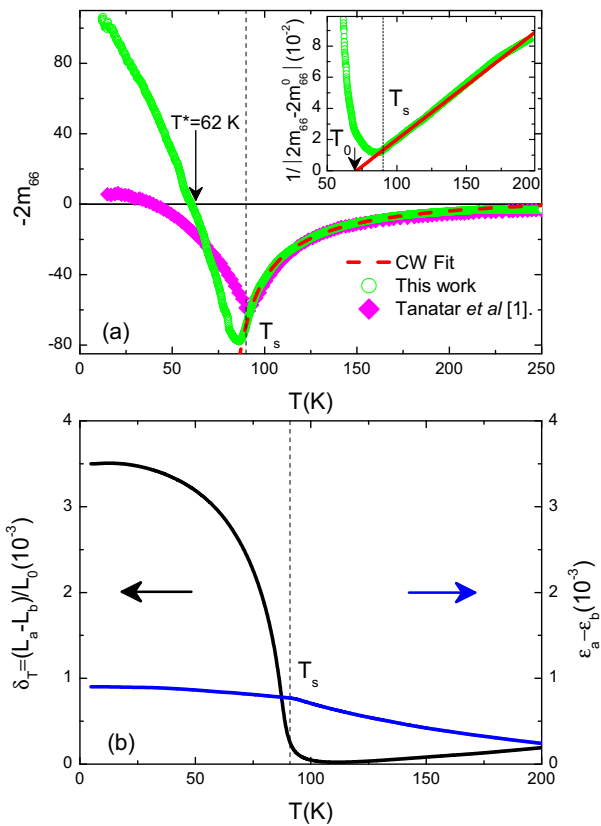

FIG. S4. (a) Calculated elastoresistivity coefficient $-2m_{66}$ (green circles) scaled with that measured by piezoelectric stack experiments (magenta squares) [1]. The red dashed line is the Curie-Weiss fitting in the form of $-2m_{66} = -2m_{66}^0 - a/(T-T_0)$ with $T_0 = 71\pm0.5$ K (Inset: inverse version of CW fitting). (b) Temperature dependence of the estimated strain $\varepsilon_a(T) - \varepsilon_b(T)$ experenced by the sample in comparison with the lattice distortion $\delta_T$ of a free standing FeSe sample.

Tanatar et al. [1] but agrees quite well with that found by Watson et al. [12], which has been attributed to strong anisotropic scattering in the orthorhombic phase [12]. We note that the magnitude of $|2m_{66}|$ at low temperature is relatively larger than that measured by Tanatar et al [1]. This is possibly due to the extremely small residual resistivity $\rho_T$ at low temperature (see Fig. S3(a) lower right inset) which is used in the denominator of Eq.(1). The estimated strain $\varepsilon_a(T) - \varepsilon_b(T)$ is plotted in Fig. S4(b) together with the sample lattice distortion, which gives a strain of $\sim 1\times 10^{-3}$ ($\sim 30\%$ of the spontaneous lattice distortion) at $T_S$.

## IV. FIELD INDUCED DETWINNING

As shown in the main text, applying field along the $[110]_{tet}$ direction effectively detwins the sample. Hence, magnetic field acts equivalently to uniaxial pressure which is commonly used in conventional detwinning devices. From the magnetostriction measurements $\Delta L_i(H)/L_0$, we can estimate the magnetic field and the corresponding pressure necessary to fully detwin $BaFe_2As_2$ and FeSe. The field induced length change $\Delta L_i(H)/L_0$ is related to the uniaxial pressure dependence of the magnetic susceptibility $\frac{d\chi_i}{dp_i}$ according to the thermodynamic Maxwell equation:

$$\lambda = \frac{1}{\mu_0 V}\frac{\partial V}{\partial H}\bigg|_{P,T} = -\frac{1}{V}\frac{\partial M}{\partial p}\bigg|_{T,H} = -\frac{1}{V}\frac{\partial \chi}{\partial p}H.$$

$$\implies \Delta L_i(H)/L_0 = -(\frac{\partial \chi_i}{\partial p_i})\mu_0 H^2. \qquad (4)$$

The quadratic behavior $\Delta L_i(H)/L_0 = c_i H^2$ is evident in the data shown in Fig.1 of the main text. Therefore the field induced distortion is also quadratic as a function of magnetic field

$$\Delta\delta_H = (\Delta L_H^{H\perp L} - \Delta L_H^{H\parallel L})/L_0 = \kappa H^2, \qquad (5)$$

where $k$ is a constant. Since the field induced distortion $\Delta\delta_H$ at field of 10 T is $\sim 10\%$ and $\sim 1\%$ of the spontaneous lattice distortion $\delta_T$ in $BaFe_2As_2$ and FeSe, respectively(see Fig.4 in the main text), the field $H_{detwin}$ needed to fully detwin the sample is given by:

$$H_{detwin} = \sqrt{\frac{\delta_T}{\Delta\delta_{H=10T}}} \times 10\,T. \qquad (6)$$

We obtain $H_{detwin} \sim 30\,T$ for $BaFe_2As_2$ which agrees excellently with earlier reports[13, 14]. For FeSe, much larger field $H_{detwin} \sim 100\,T$ is necessary as the susceptibility anisotropy is fairly weak.

From magnetostriction data presented in the Fig.1 of main text and according to Eq.4, the uniaxial pressure dependence of the magnetic susceptibility of $BaFe_2As_2$ is obtained $\frac{d\chi_b}{dp_b} \sim 1.5\times 10^{-6}\,MPa^{-1}$ at 100 K. In the twinned state, the system has susceptibility $\chi_0(H, P = 0) = \frac{\chi_b + \chi_a}{2}$. For $BaFe_2As_2$, the susceptibility changes to $\chi_b$ in the fully detwinned state with one single domain where the shorter axis $b$ aligns in the field direction. Then the pressure required to fully detwin $BaFe_2As_2$ is:

$$P_{detwin} = \frac{(\chi_b - \chi_a)}{2\cdot d\chi_b/dp_b} \qquad (7)$$

The susceptibility anisotropy of $BaFe_2As_2$ at 100 K is $\chi_b - \chi_a \sim 3\times 10^{-5}$, and we have $P_{detwin} \sim 10\,MPa$ at 100 K which matches well with the value obtained by applying uniaxial pressure directly[?]. For FeSe, $\frac{d\chi_a}{dp_a} \sim 2.5\times 10^{-7}\,MPa^{-1}$ and $\chi_a - \chi_b \sim 7.5\times 10^{-6}$ at 10 K, hence $P_{detwin} \sim 15\,MPa$. The effective field induced pressure at 10 T is then $P_{H=10T} \sim 3\,MPa$ and $P_{H=10T} \sim 0.15\,MPa$ for $BaFe_2As_2$ and FeSe, respectively.

---

* christoph.meingast@kit.edu